\documentclass[a4paper,11pt]{article}
\usepackage{pos}

\usepackage{graphicx}
\usepackage{epstopdf}
\usepackage{cancel}

\newcommand{\beq}{\begin{equation}}
\newcommand{\eeq}{\end{equation}}
\newcommand{\bea}{\begin{eqnarray}}
\newcommand{\eea}{\end{eqnarray}}

\newcommand{\eq}{Eq.~}
\newcommand{\eqs}{Eqs.~}
\newcommand{\fig}{Fig.~}
\newcommand{\figs}{Figs.~}

\newcommand{\tr}{{\rm Tr}}
\newcommand{\bx}{{\bf x}}

\newcommand{\bp}{{\bf p}}

\def\lsi{\raise0.3ex\hbox{$<$\kern-0.75em\raise-1.1ex\hbox{$\sim$}}}
\def\gsi{\raise0.3ex\hbox{$>$\kern-0.75em\raise-1.1ex\hbox{$\sim$}}}

\title{On chiral spin symmetry and the QCD phase diagram}
\ShortTitle{Chiral spin symmetry}

\author*[a,b]{Owe Philipsen}
\author[c]{Leonid Ya. Glozman}
\author[a]{Peter Lowdon}
\author[d]{Robert D. Pisarski}

\affiliation[a]{Institute for Theoretical Physics, Goethe University Frankfurt,\\
 Max-von-Laue-Str.\ 1, D-60438 Frankfurt am Main, Germany}
 \affiliation[b]{John von Neumann Institute for Computing (NIC)\\
at GSI, Planckstr.\ 1, 64291 Darmstadt, Germany}
\affiliation[c]{Institute of Physics, University of Graz, A-8010 Graz, Austria}
\affiliation[d]{Physics Department, Brookhaven National Laboratory, Upton, NY 11973, USA}

\emailAdd{philipsen@itp.uni-frankfurt.de}

\abstract{
Recently, an approximate $SU(4)$ chiral spin-flavour symmetry was observed in multiplet patterns
of QCD meson correlation functions, in a temperature range above the chiral crossover.
This symmetry is larger than the chiral symmetry of massless QCD, and can only arise effectively 
when colour-electric quark-gluon interactions dynamically dominate the quantum effective action.
At temperatures about three times the crossover temperature,  these patterns disappear again, 
indicating the screening of colour-electric interactions, and the expected chiral symmetry is recovered.
In this contribution we collect independent evidence for such an intermediate temperature range, 
based on screening masses and the pion spectral function. Both kinds of observables behave non-perturbatively in this 
window, with resonance-like peaks for the pion and its first excitation disappearing gradually with temperature.
Using symmetry arguments and the known behaviour of screening masses at small densities, we discuss
how this chiral spin symmetric band continues into the QCD phase diagram.   
}

\FullConference{%
The 39th International Symposium on Lattice Field Theory,\\
8th-13th August, 2022,\\
Rheinische Friedrich-Wilhelms-Universität Bonn, Bonn, Germany
}

\begin{document}
\maketitle
	
\section{Introduction}

While QCD at
finite baryon density cannot be simulated by Monte Carlo methods because of a severe sign problem, 
the physics at finite temperature poses no technical complications and, given sufficient compute power, can 
be evaluated non-perturbatively. In particular, the nature of the chiral transition as an analytic crossover \cite{Aoki:2006we} and 
its associated pseudo-critical temperature $T_\mathrm{pc}$ \cite{HotQCD:2018pds,Borsanyi:2020fev} 
are known with ever increasing precision. 

However, the physics at temperatures above the chiral crossover appears to be different and more complex than expected so far. 
A few years ago  an additional chiral spin symmetry was observed to emerge dynamically in a temperature range of 
roughly $T_\mathrm{pc}\lsi T \lsi 3T_\mathrm{pc}$ \cite{Rohrhofer:2019qwq,Rohrhofer:2019qal}. 
This symmetry is larger than the chiral symmetry and can only be 
realised approximately if colour-electric quark-gluon interactions dominate the quantum effective action. 
This suggests that there are three temperature regimes in QCD with different effectively realised symmetries and dynamics.
In this contribution we review the original evidence based on multiplets of spatial and temporal 
correlators \cite{Rohrhofer:2019qwq,Rohrhofer:2019qal}, and show that the associated change of dynamics
is also visible in screening masses and the pion spectral function. We then discuss how this new band of 
effectively chiral spin symmetric QCD extends into the QCD phase diagram.  
	
\section{Emergent chiral spin symmetry at finite $T$}

\begin{figure}[t]
\vspace*{-0.5cm}
\centering
\includegraphics[width=0.32\textwidth]{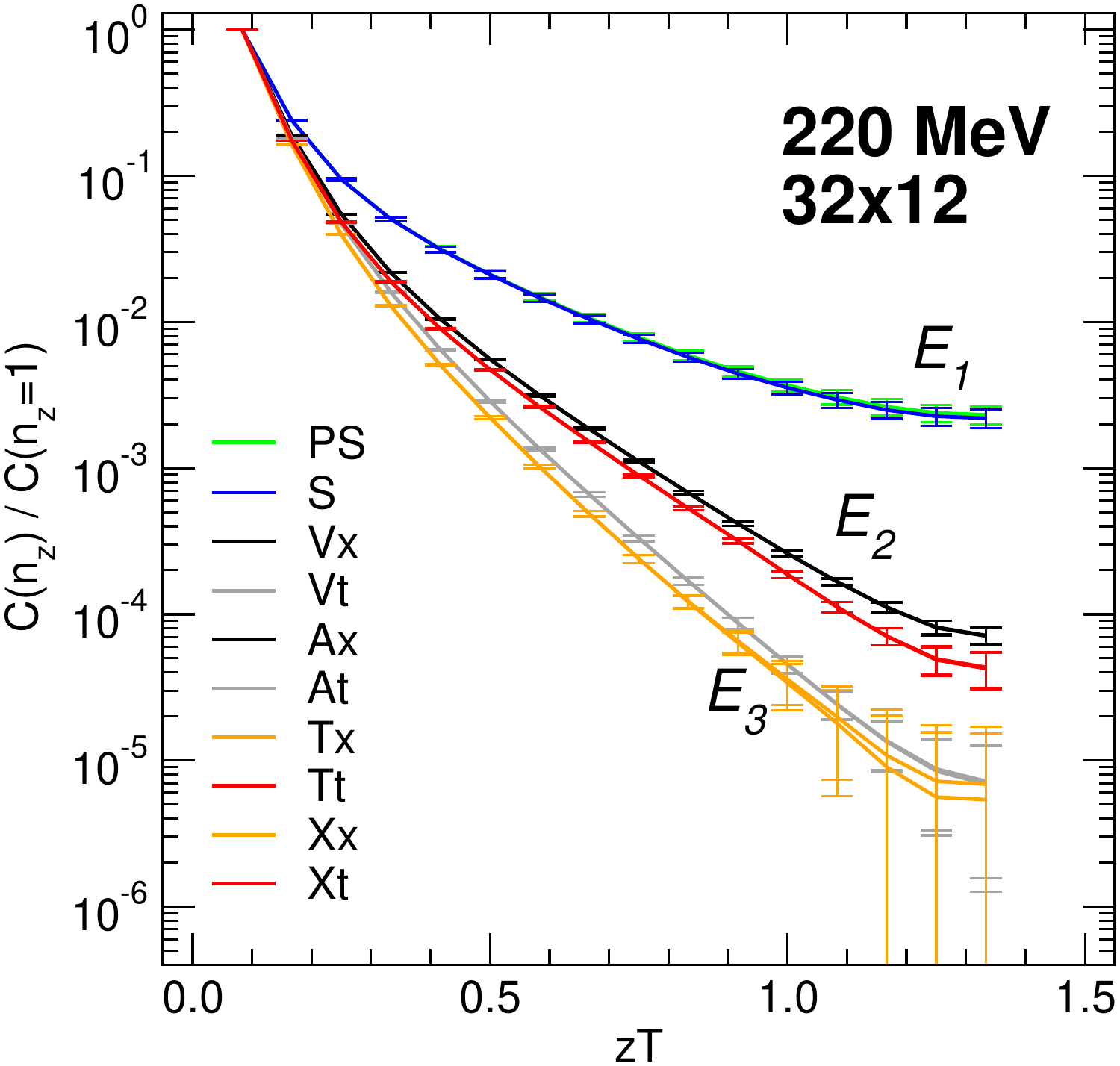}
\includegraphics[width=0.32\textwidth]{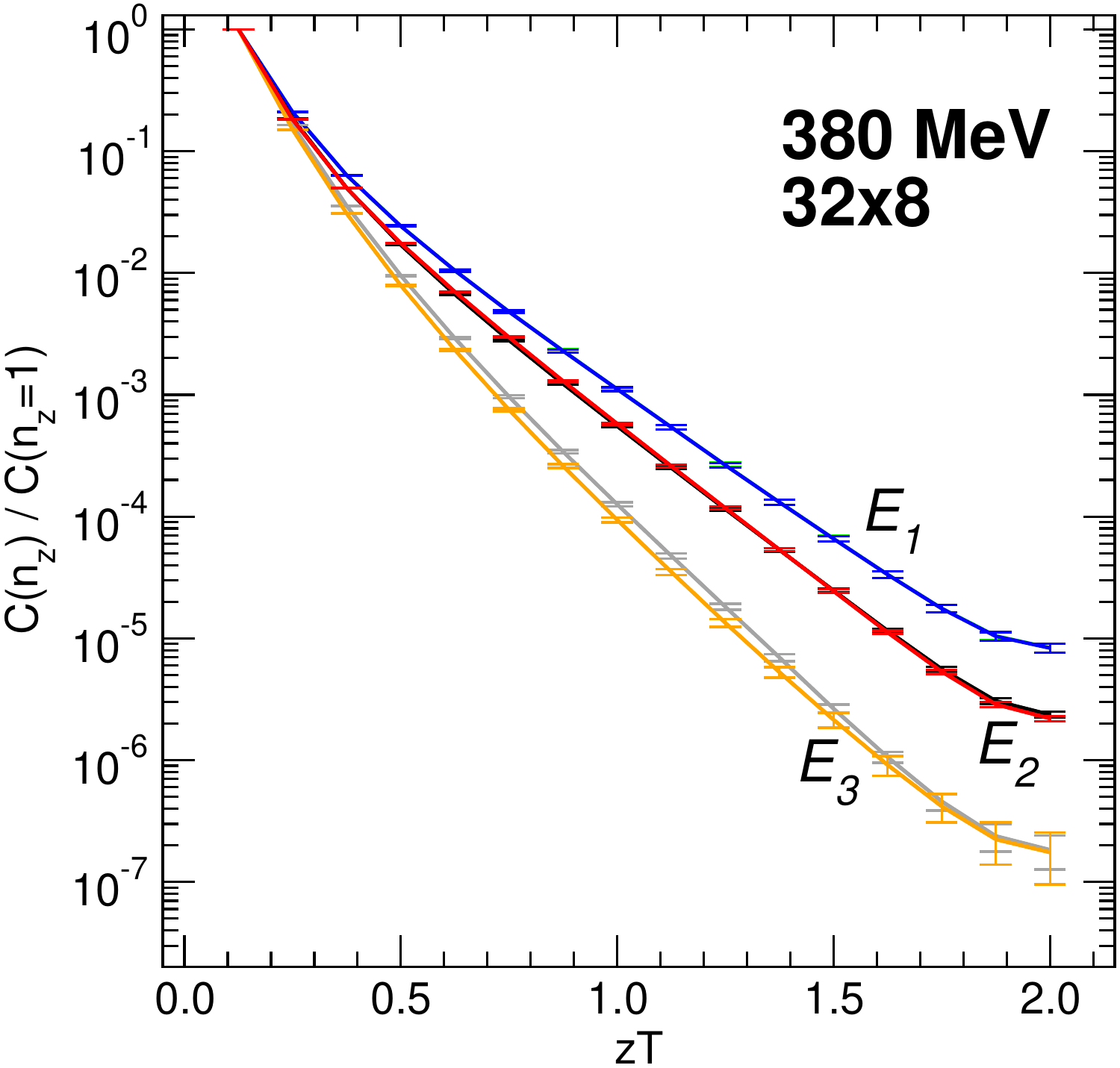}
\includegraphics[width=0.32\textwidth]{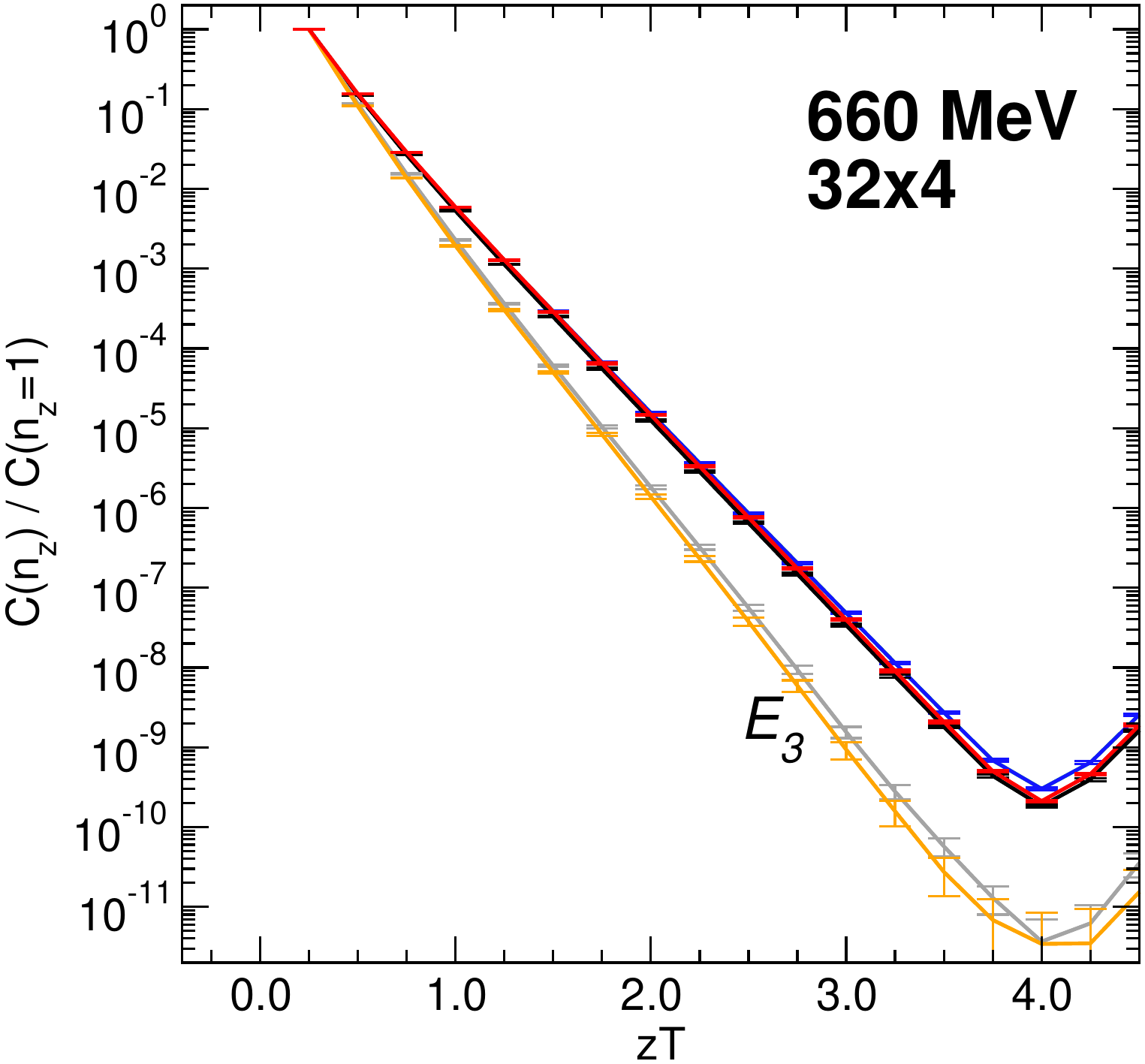}
\caption[]{Spatial correlation functions show distinct $E_1,E_2,E_3$ multiplets of the approximate $SU(4)$ chiral spin symmetry, at 
temperatures above the crossover. 
At large temperatures, these reduce to the multiplets of the ordinary chiral symmetry. Based on domain wall 
fermions, from \cite{Rohrhofer:2019qwq}.}
\label{fig:chi_spin1}
\end{figure}

Consider a $SU(2)_{CS}$ chiral spin transformation of Dirac quark fields defined by
\beq
\psi(x)\rightarrow \exp\Big( i\vec{\Sigma} \cdot\vec{\epsilon}\Big)\psi(x)\;,
\quad \vec{\Sigma}=(\gamma_k,-i\gamma_5\gamma_k,\gamma_5)\;,
\quad [\Sigma_i,\Sigma_j]=2i \epsilon_{ijk}\Sigma_k\;.
\eeq   
Here $k=0,\ldots 3$ can be any of the euclidean gamma matrices, and it is easy to check that the generators $\vec{\Sigma}$
satisfy a $SU(2)$ algebra. It is apparent that $SU(2)_{CS}\supset U(1)_A$. Furthermore, when combined with 
isospin, $SU(2)_{CS}\otimes SU(2)_V$ can be embedded into the larger $SU(4)$, 
which contains the usual chiral symmetry of the massless QCD Lagrangian, 
$SU(4)\supset SU(2)_L\times SU(2)_R\times U(1)_A$.

The QCD Lagrangian is not invariant under chiral spin transformations. However, a thermal medium implies a preferred
Lorentz frame, and the massless quark action can be written as 
\begin{equation}
\bar{\psi}\gamma_\mu D_\mu \psi= \bar{\psi}\gamma_0 D_0 \psi + \bar{\psi}\gamma_i D_i\psi\;, \quad\mbox{with}\quad
[\Sigma_i,\gamma_0\gamma_0]=0\;, \;\;[\Sigma_i,\gamma_0\gamma_j]\neq 0\;.
\label{eq:comm}
\end{equation}
One finds the colour-electric part of the quark-gluon interaction to 
be CS- and $SU(4)$-invariant, while kinetic terms (and thus the free Dirac action) and colour-magnetic interactions are not. 
Hence, chiral spin symmetry is never exact in physical QCD, but its approximate realisation is possible
if the colour-electric quark-gluon interaction dominates the quantum effective action in some dynamical range. 

On the lattice, symmetries are 
straightforwardly tested by investigating degeneracy patterns in correlation functions. Consider the euclidean meson correlators
 with $J=0,1$ and $\Gamma$ some appropriate Dirac matrix,
\begin{equation}
C_\Gamma(\tau,\bx)=\langle O_\Gamma(\tau,\bx)\,O_\Gamma^\dag(0,\mathbf{0})\rangle\;.
\end{equation}
These carry the full information about all excitations in their
associated spectral functions $\rho_\Gamma(\omega,\bp)$,
\begin{eqnarray}
C_\Gamma(\tau,\bp) =\int_0^\infty \frac{d\omega}{2\pi}\;K(\tau,\omega)\rho_\Gamma(\omega,\bp)\;, \quad
K(\tau,\omega)=\frac{\cosh(\omega(\tau-1/2T))}{\sinh(\omega/2T)}\;.
\label{eq:corr}
\end{eqnarray}
For an isotropic system in equilibrium, it is sufficient to probe 
the spatial and temporal correlators averaged over the orthogonal directions,
\begin{eqnarray}
C_\Gamma^s(z)=\sum_{x,y,\tau} C_\Gamma(\tau,\bx)\;,\label{eq:c_z}\;\quad
C_\Gamma^\tau(\tau)=\sum_{x,y,z}C_\Gamma(\tau,\bx)\;.
\end{eqnarray}
\begin{figure}[t]
\vspace*{-0.5cm}
\centering
\includegraphics[width=0.4\textwidth]{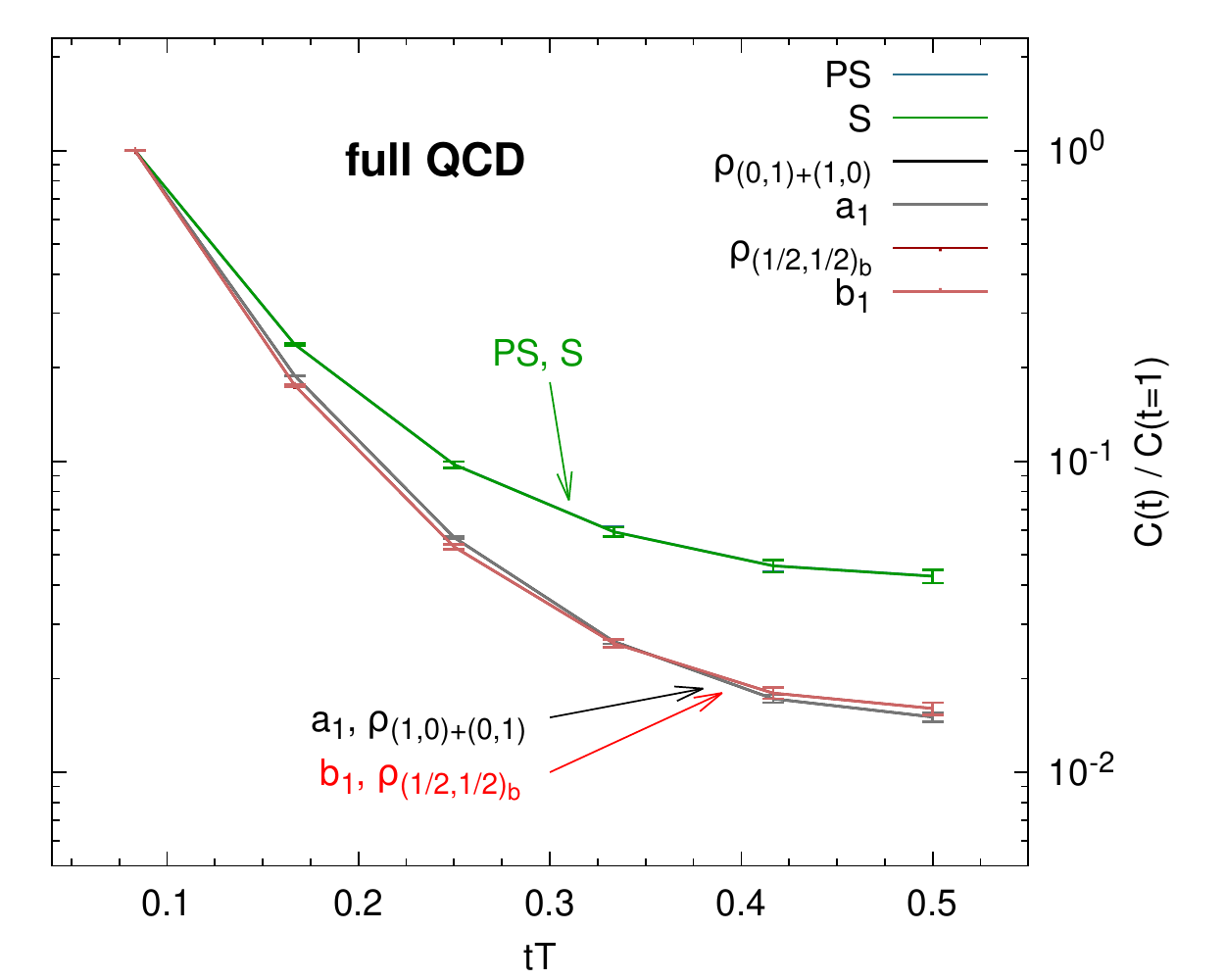}
\includegraphics[width=0.4\textwidth]{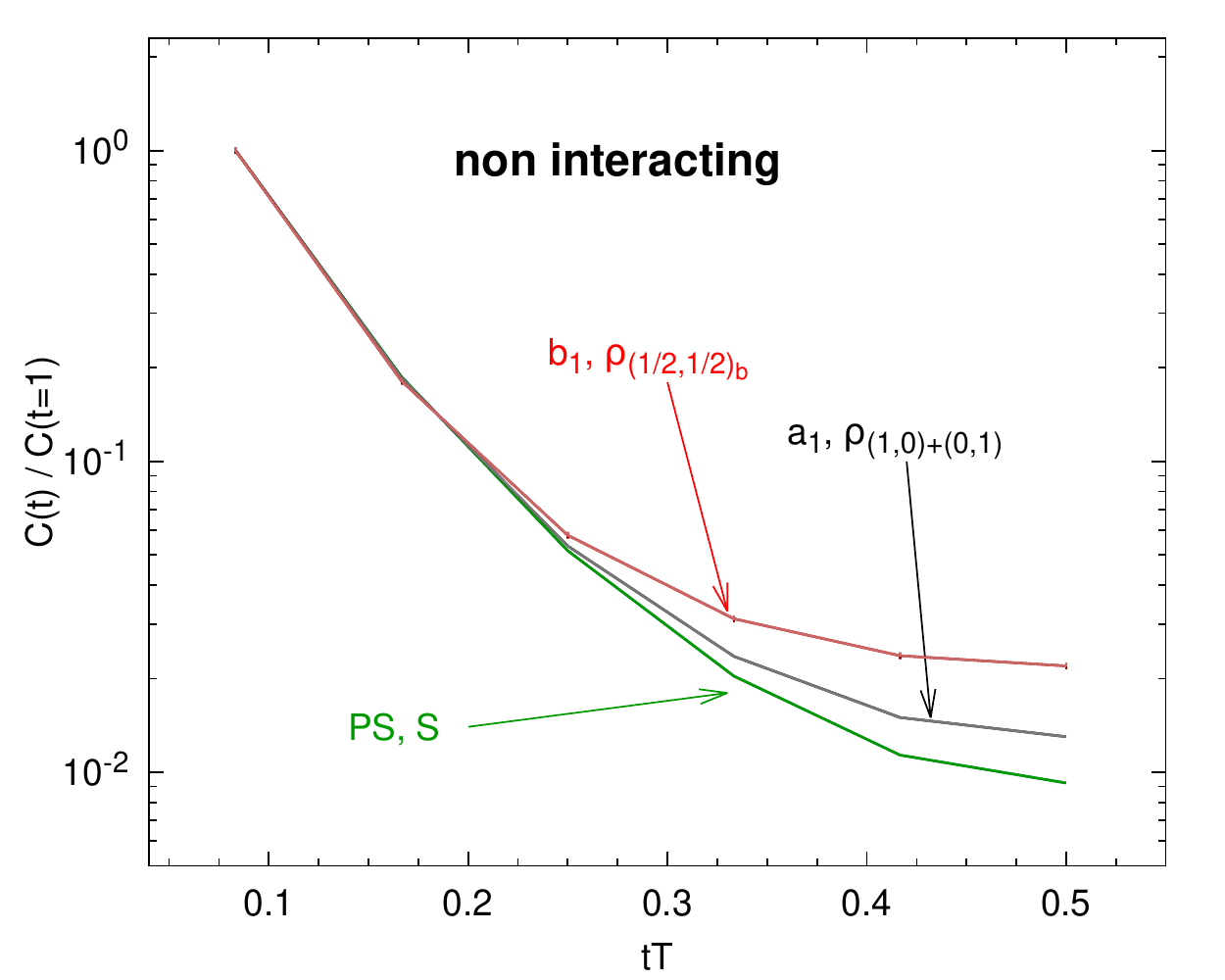}
\caption[]{
Temporal correlation functions on $12 \times 48^3$ lattices. 
    Left: Full QCD results at $T=220$ MeV, representing multiplets of all groups, $U(1)_A$, 
    $SU(2)_L \times SU(2)_R$, $SU(2)_{CS}$ and $SU(4)$.
    Right: Correlators calculated with free quarks with manifest $U(1)_A$ and $SU(2)_L \times SU(2)_R$ symmetries.
   From \cite{Rohrhofer:2019qal}.}
\label{fig:chi_spin2}
\end{figure}

Numerical results obtained with $N_f=2$ JLQCD domain wall fermions with good chiral symmetry and 
all lattice spacings $<0.1$ fm \cite{Rohrhofer:2019qwq,Rohrhofer:2019qal}
are shown in \figs\ref{fig:chi_spin1} and \ref{fig:chi_spin2}, respectively. \fig\ref{fig:chi_spin1} shows three
multiplets of spatial correlators, $E_{1,2,3}$, at different temperatures. Of these,
$E_1$ is due to $U(1)_A$ restoration whereas $E_3$ requires the full chiral symmetry. Both multiplets are expected
above the chiral crossover. What is surprising is the multiplet $E_2$, which does \textit{not} correspond to 
a representation of chiral symmetry, but to one of the larger $SU(4)$ (for the representations and their identification 
with meson states, see \cite{Glozman:2014mka,Glozman:2015qva}). Its distinctive appearance demonstrates the
dynamical emergence of chiral spin symmetry in this regime. As temperature is further increased, $E_2$ gradually 
disappears as a separate multiplet and only those belonging to the expected chiral symmetry survive. 

This finding is confirmed by the temporal correlators shown in \fig\ref{fig:chi_spin2} (left). In this case it is the degeneracy of 
the $a_1$ and $b_1$ correlators, which cannot be transformed into each other by ordinary chiral transformations, but are
related via the larger $SU(4)$ symmetry. By contrast, \fig\ref{fig:chi_spin2} (right) shows the same correlators evaluated 
with free quarks, corresponding to the leading perturbative result, which respects the ordinary chiral symmetry. The degeneracy
pattern and ordering are qualitatively incompatible with the QCD data. 

One must then conclude that there are three temperature
ranges with different symmetry properties: the regime of broken chiral symmetry at low temperatures, 
an approximately $SU(4)$-symmetric regime
for   $T_\mathrm{pc}\lsi T \lsi 3T_\mathrm{pc}$, and the expected chirally symmetric regime at high temperatures. Usually
changes in symmetry imply changes in dynamical behaviour and effective degrees of freedom. We now show that such
changes are indeed visible also in other observables.

\section{Screening masses \label{sec:masses}}

Although not directly measurable experimentally, QCD screening masses are valuable and well-studied observables, 
because they
can be readily evaluated both non-perturbatively and pertubratively, permitting insights into the transition between the 
confined and deconfined regimes. They can be extracted from the exponential decay of spatial correlators,
\beq
C_\Gamma^s(z)\stackrel{z\rightarrow \infty}{\longrightarrow} \mbox{const.}\;e^{-m_{scr}z}\;,
\eeq  
and correspond to the ``zero momentum'' eigenvalues of a ''spatial Hamiltonian'' $H_z$, which acts on 
states defined over $\{x,y,\tau\}$-space and translates
them in $z$ direction,
\beq
|\psi(\tau,x,y;,z+1)\rangle=e^{-aH_z}|\psi(\tau,x,y;z)\rangle\;.
\eeq

\begin{figure}[t]
    \centering
    \begin{minipage}{0.48\textwidth}
       \centering
      \vspace*{-0.2cm}
       \includegraphics[width=\linewidth]{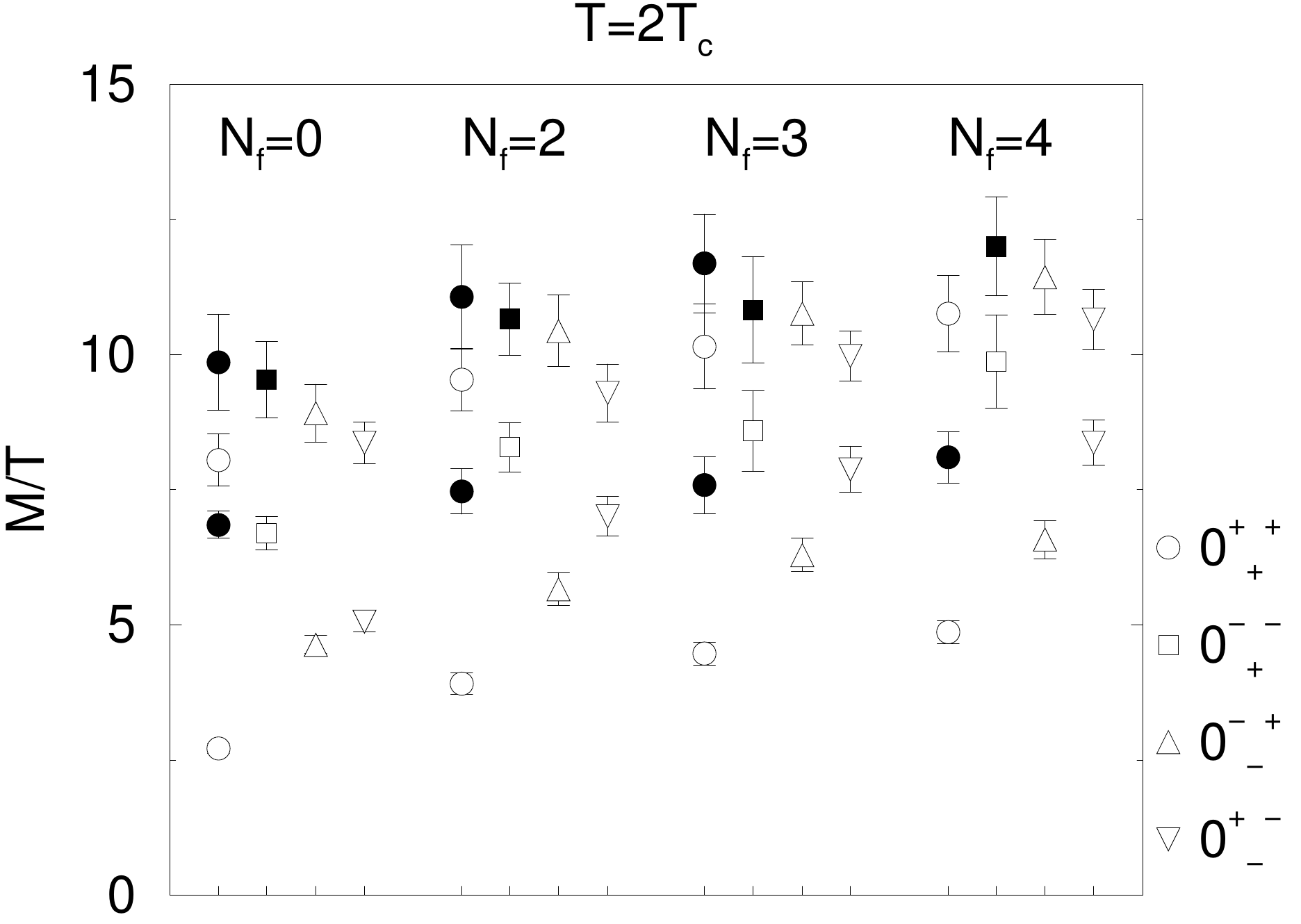}
       \captionof{figure}{
              Screening masses computed within EQCD, after integrating out $N_f$ massless fermions.
               Filled symbols correspond to operators constructed from colour-magnetic fields exclusively, $\tr{F_{ij}^2},\ldots$,
                open symbols contain colour-electric fields $\tr(A_0^2),\tr(A_0F_{ij}),\ldots$. 
                From \cite{Hart:2000ha}. 
            }\label{fig:eqcd}  
     \end{minipage}%
     \hfill
     \begin{minipage}{0.42\textwidth}
       \centering
       \vspace*{-0.5cm}
       \includegraphics[width=\linewidth]{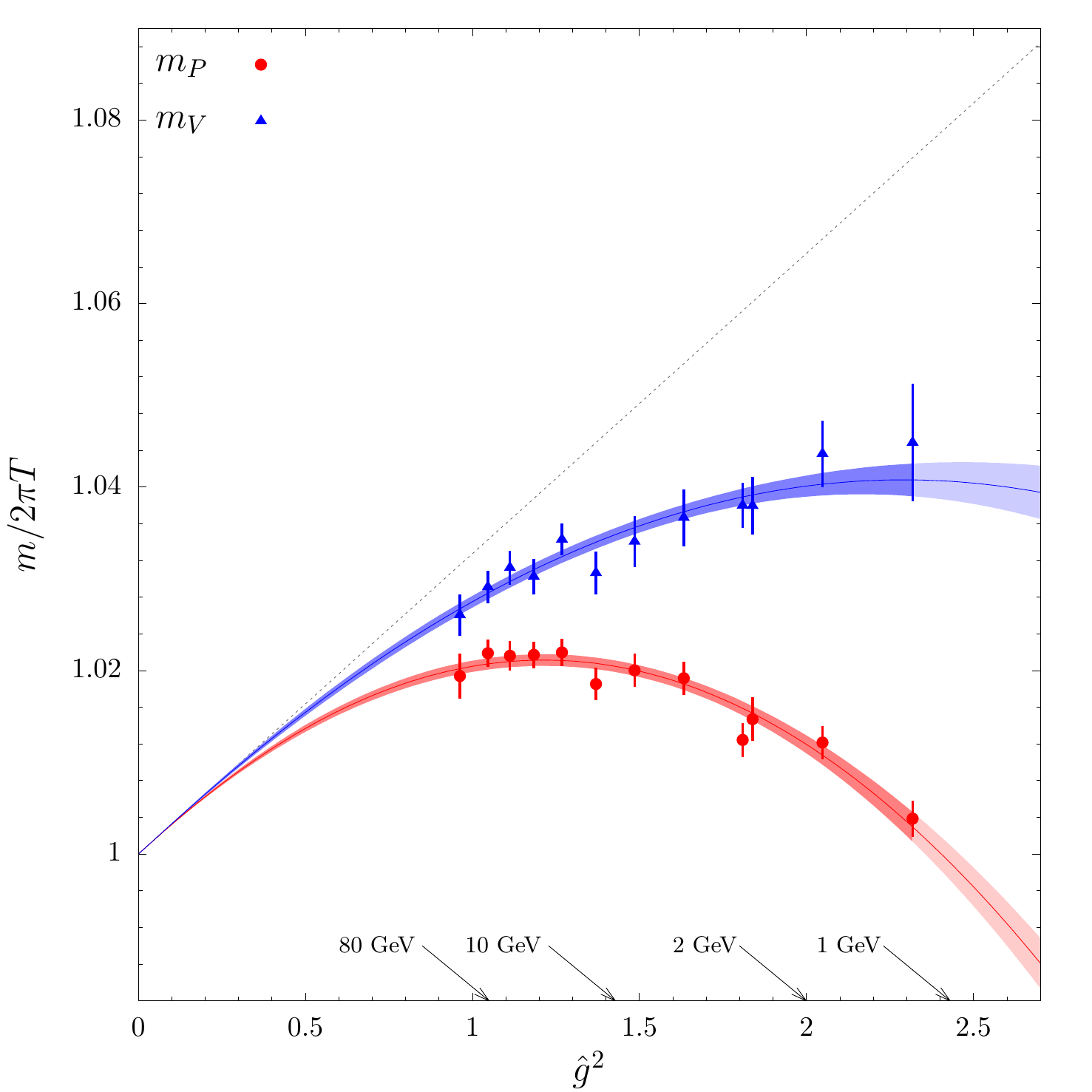}
       \captionof{figure}{
            Pseudo-scalar and vector screening masses at high temperatures, based on $O(a)$-improved 
            Wilson fermions with modified boundary conditions and step scaling techniques. From \cite{DallaBrida:2021ddx}.
            }\label{fig:hight}  
     \end{minipage}
\end{figure}

To begin, we recall evidence for the dynamical dominance of the colour-electric fields, which was produced 
in the context of dimensional reduction of finite temperature quantum field theories. It is well known that finite temperature
gauge theories are characterised by three dynamical scales, the hard scale $\sim \pi T$ for the non-zero Matsubara/fermion modes,
the soft scale $\sim gT$ for the colour-electric fields and the ultra-soft scale $\sim g^2T$ for the entirely non-perturbative magnetic modes. 
In \cite{Hart:2000ha} EQCD, obtained as an effective bosonic theory after integrating out the hard modes perturbatively, 
was simulated and screening masses were computed for a range of quantum numbers,
as shown in \fig\ref{fig:eqcd}. One observes that, independent of the quantum number channel, the screening masses
based on interpolating operators containing at least one $A_0$ field are considerably smaller than those constructed entirely
from $A_i$-fields, which means the dynamics is dominated by colour-electric fields. This is reversed compared to the perturbative
parametric ordering, and provides the necessary condition for an approximate realisation of chiral spin symmetry.

Next, we consider meson correlators in QCD, i.e.~screening states living on the hard scale, which is the most likely to behave 
perturbatively.  Recently, by employing shifted boundary conditions in combination with step scaling techniques on $N_f=2$ 
$O(a)$-improved Wilson fermions, it was
possible to compute screening masses on the lattice in the high temperature regime, $T\sim 1-160$~GeV for the fist time,
with unprecedented precision \cite{DallaBrida:2021ddx}. Results for the pseudo-scalar and vector screening masses are shown 
in  \fig\ref{fig:hight}, the coloured bands represent fits of the data to the perturbative parametrisation
\begin{eqnarray}
\frac{m_{PS}}{2\pi T}=1+p_2 \,\hat{g}^2(T) + p_3 \, \hat{g}^3(T)+p_4 \,\hat{g}^4(T)\;,\qquad 
\frac{m_{V}}{2\pi T}=\frac{m_{PS}}{2\pi T} + s_4\, \hat{g}^4(T)\;.
\label{eq:psv}
\end{eqnarray}
Here $\hat{g}^2(T)$ denotes the running coupling renormalised in the $\overline{\mathrm{MS}}$-scheme 
at $\mu=2\pi T$, while $p_2\ldots p_4,s_4$ are numbers. 
The perturbative value of $p_2$ \cite{Laine:2003bd} is fully confirmed, as is the corrsponding functional form
\eq(\ref{eq:psv}), over a remarkable three orders of magnitude in temperature. The difference due to spin can be described
by a single $\hat{g}^4$-term. 

\begin{figure}[t]
\vspace*{-0.5cm}
\centering
\includegraphics[width=0.45\textwidth]{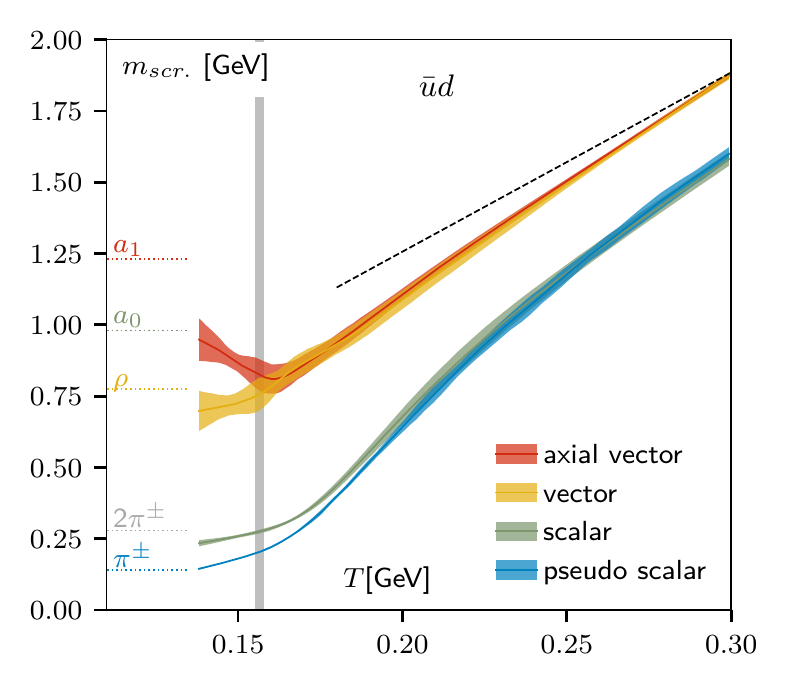}
\includegraphics[width=0.45\textwidth]{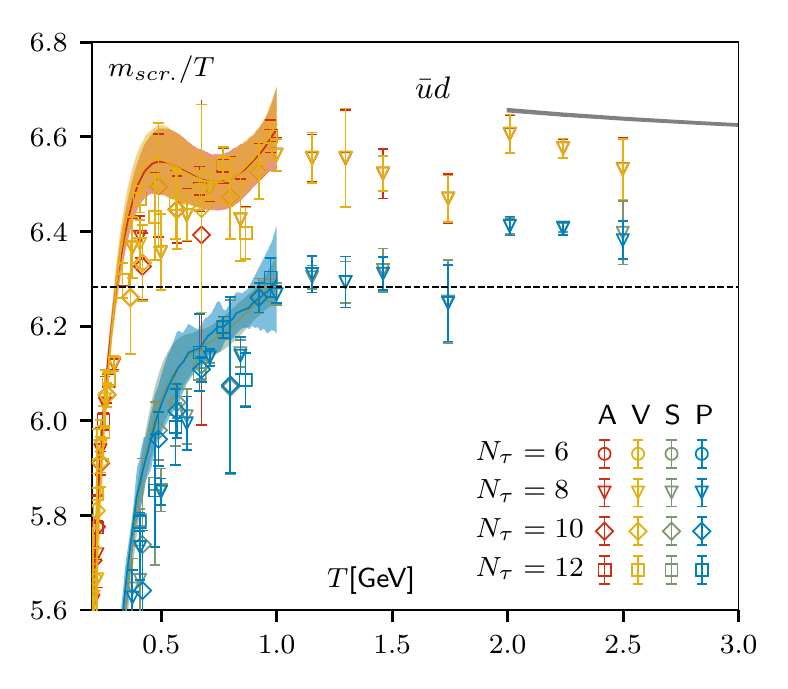}
\caption[]{
Screening masses of the lightest $\bar{u}d$-mesons, evaluated using 
    HISQ fermions. From \cite{Bazavov:2019www}.}
\label{fig:mscr}
\end{figure}

Two dynamical changes, incompatible with perturbative behaviour, become apparent in the intermediate range 
$T\sim~0.15-1$ GeV, which was studied
with $N_f=2+1$ HISQ fermions at physical quark masses \cite{Bazavov:2019www}, as shown in \fig\ref{fig:mscr}. On the left
one observes chiral symmetry restoration, with screening masses becoming degenerate at $T_\mathrm{pc}\approx 150$ MeV, as expected (the degeneracy of
the pion with the two-pion state instead of the $a_0$ is a known artefact of the staggered taste 
splitting~\cite{Prelovsek:2005rf,Bazavov:2019www}). 
More interestingly, another marked change of dynamics is visible in the right plot, where $m_{scr}/T$ changes from nearly vertical to
almost horizontal behaviour within a narrow temperature range. At its high temperature end, the plot matches the behaviour
from the previous figure, with all screening masses overshooting $\sim 2\pi T$ and a sizeable spin effect.
Nevertheless (and ignoring the wiggles within errors), the horizontal part of the plot is fully compatible with \eq(\ref{eq:psv})
and its logarithmic $T$-dependence, i.e.~perturbative behaviour. This changes abruptly between $T\sim 500-700$ MeV for
all four quantum number channels shown in \fig\ref{fig:mscr}. The sudden change in $T$-dependence is not 
compatible with the perturbative behaviour \eq(\ref{eq:psv}), nor can it be accomodated by higher order corrections. 
The same abrupt bending is observed in the same $T$-range also for the $\bar{u}s$ 
and $\bar{s}s$ mesons~\cite{Bazavov:2019www}, 
i.e.~across 12 different quantum number channels, and therefore suggests a change of dynamics for the entire system. 
The temperature region, where this dynamical change happens, 
is fully compatible with the upper boundary of chiral spin symmetry, which 
explains the breakdown of perturbation theory: since the latter is organised around free quarks, it cannot accomodate
chiral spin symmetry to any finite order.  This breakdown of perturbation theory confirms the earlier conclusion based on symmetry
alone, i.e.~the presence of non-perturbative colour-electric interactions between quarks. For this reason this regime has 
been termed ''stringy fluid''  \cite{Rohrhofer:2019qwq,Rohrhofer:2019qal}.

\section{The pion spectral function}

\begin{figure}[t]
\vspace*{-0.5cm}
\centering
\includegraphics[width=0.45\textwidth]{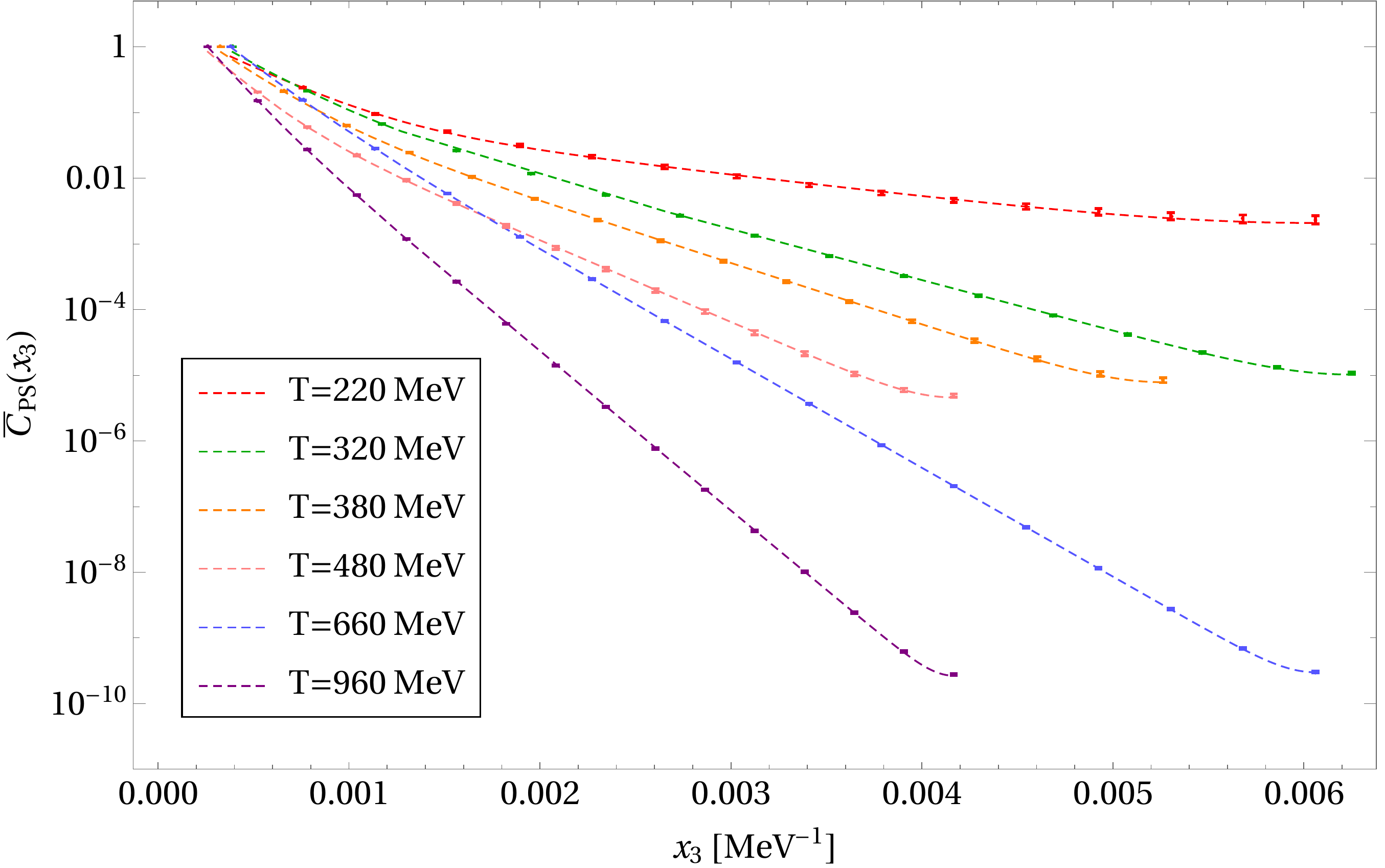}
\includegraphics[width=0.5\textwidth]{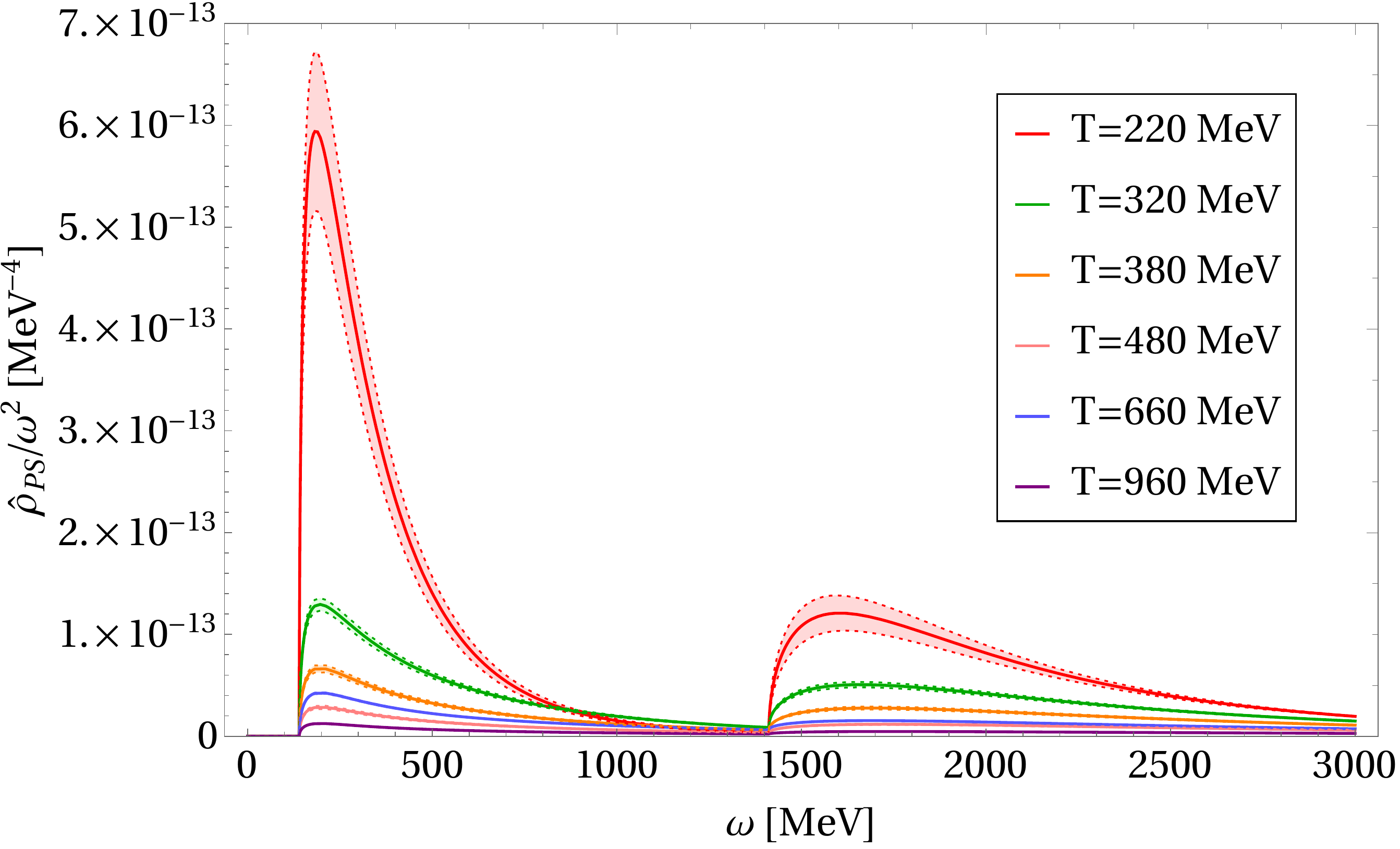}
\caption[]{Left: Pseudo-scalar, spatial correlation functions  from \fig\ref{fig:chi_spin1} \cite{Rohrhofer:2019qwq}, the lines
represent two-state exponential fits. Right: Spectral function resulting from those fits and the ansatz \eqs(\ref{eq:decomp},\ref{C_int}).}
\label{fig:spec}
\end{figure}

More direct information about the nature of the effective degrees of freedom in the different regimes may be expected from the 
spectral functions, \eq(\ref{eq:corr}). Unfortunately, their extraction from discrete sets of lattice correlator data represents
an ill-posed inversion problem. Here we try a new method  that applies to stable scalar particles in a heat bath, i.e.~to the pion 
in the case of QCD, which allows to circumvent the integral inversion.

The method is based on the fundamental principle of locality (in the sense of micro-causality)
of quantum field theories. It ensures a representation of the spectral function~\cite{Bros:1998ua,Bros:1996mw} as
\begin{align}
\rho_{\text{PS}}(\omega,\bp) = \int_{0}^{\infty} \! ds \int \! \frac{d^{3}u}{(2\pi)^{2}} \ \epsilon(p_{0}) \, \delta\!\left(\omega^{2} - (\bp-\mathbf{u})^{2} - s \right)\widetilde{D}_{\beta}(\mathbf{u},s)\;,
\label{eq:commutator_rep}
\end{align}
with $\beta=1/T$, the thermal spectral density $\widetilde{D}_{\beta}(\mathbf{u},s)$. Note that the standard 
K\"allen-Lehmann vacuum representation is attained smoothly as $T\rightarrow 0$. For stable massive particles, 
such as QCD pions, 
the authors argue for the analytic vacuum structure of the spectral density
to be preserved in the absence of a true phase transition, and propose an ansatz with particle and scattering contributions,  
\begin{align}
\widetilde{D}_{\beta}(\mathbf{u},s)= \widetilde{D}_{m,\beta}(\mathbf{u})\, \delta(s-m^{2}) + \widetilde{D}_{c, \beta}(\mathbf{u},s)\;.
\label{eq:decomp}
\end{align} 
In an isotropic medium the spatial correlators and the spectral density  
are then related by~\cite{Lowdon:2022xcl}
\begin{align}
C^s_{PS}(z) = \frac{1}{2}\int_{0}^{\infty} \! ds \int^{\infty}_{|z|} \! dR \ e^{-R\sqrt{s}} D_{\beta}(R,s).
\label{C_int}
\end{align}
For temperatures below the threshold to the scattering states we then expect the first term 
in \eq(\ref{eq:decomp}) to dominate. Neglecting the continuum part, the calculation of the spectral function is straightforward.
First, we fit the spatial pseudo-scalar correlators from \fig\ref{fig:chi_spin1} by the sum of two exponentials representing the $\pi,\pi^*$,
which gives an excellent description of the data in the entire temperature range, cf.~\fig\ref{fig:spec} (left). This provides
the $D_{m,\beta}(|\bx|)=\alpha_{\pi,\pi^*}\exp(-\gamma_{\pi,\pi^*} |\bx|)$, 
from which the spectral function can be reconstructed 
using~\eqs(\ref{eq:commutator_rep},\ref{eq:decomp}) and the vacuum masses $m_\pi,m_{\pi^*}$,
\begin{align}                                                   
\rho_{\text{PS}}(\omega,\bp=0)=   \epsilon(\omega) \left[ \theta(\omega^{2}-m_{\pi}^{2}) \, \frac{4\, \alpha_{\pi} \,  \gamma_{\pi} \sqrt{\omega^{2}-m_{\pi}^{2}}}{(\omega^{2}-m_{\pi}^{2}+\gamma_{\pi}^{2})^{2}} +\theta(\omega^{2}-m_{\pi^{*}}^{2}) \, \frac{4\, \alpha_{\pi^{*}} \,  \gamma_{\pi^{*}} \sqrt{\omega^{2}-m_{\pi^{*}}^{2}}}{(\omega^{2}-m_{\pi^{*}}^{2}+\gamma_{\pi^{*}}^{2})^{2}}   \right].
\label{eq:spec1}
\end{align}
The result is shown in \fig\ref{fig:spec} (right) and displays the vacuum thresholds followed by a pronounced
resonance-like peak structure for both the pion and its first excitation. As the temperature increases, the peaks widen 
and gradually disappear into a continuum, consistent with sequential hadron melting, albeit at temperatures
significantly above $T_\mathrm{pc}$.
This is in accord with the approximately chiral-spin symmetric window with non-perturbative, hadron-like
excitations. 

\begin{figure}[t]
\vspace*{-0.5cm}
\centering
\includegraphics[width=0.45\textwidth]{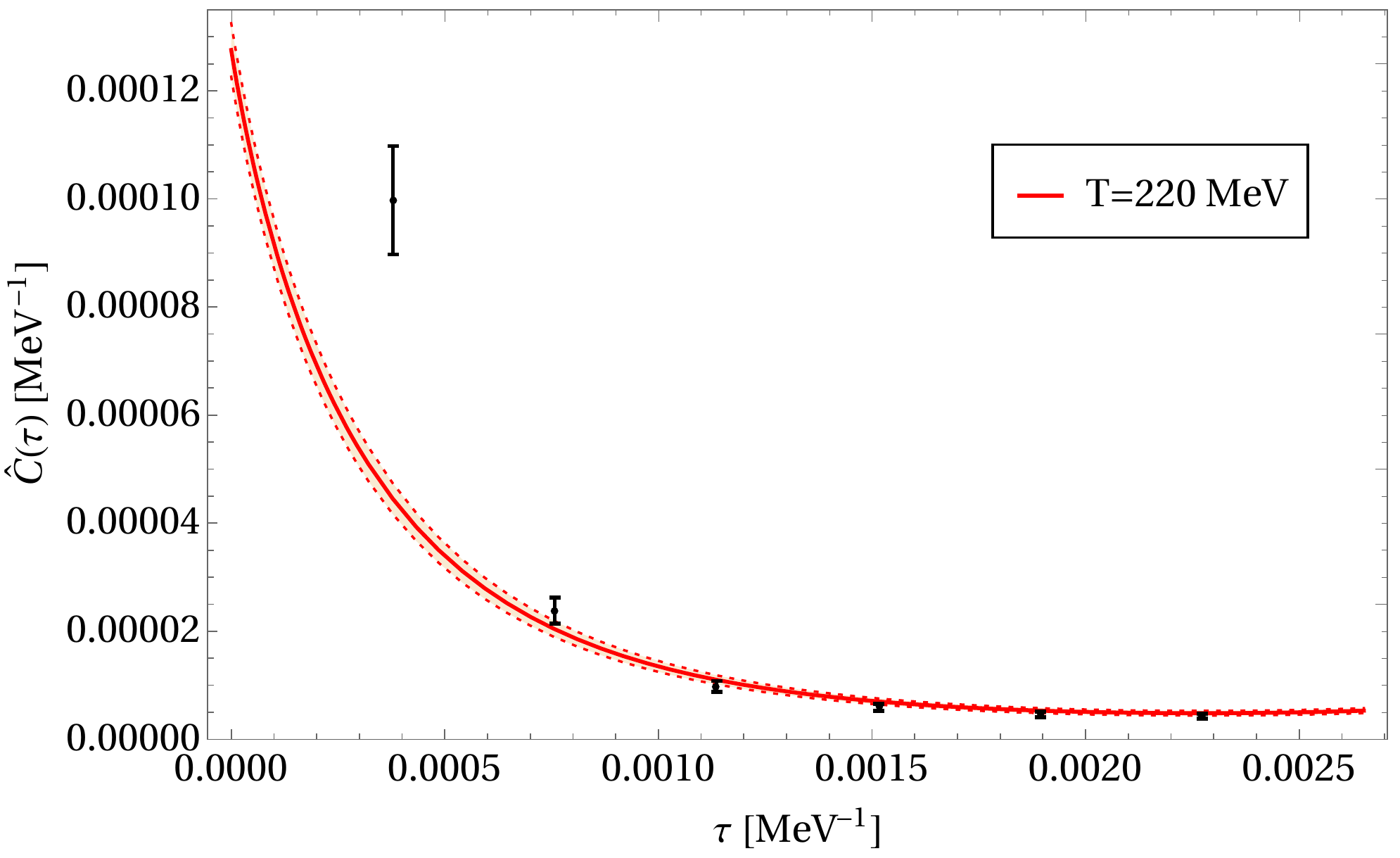}
\includegraphics[width=0.45\textwidth]{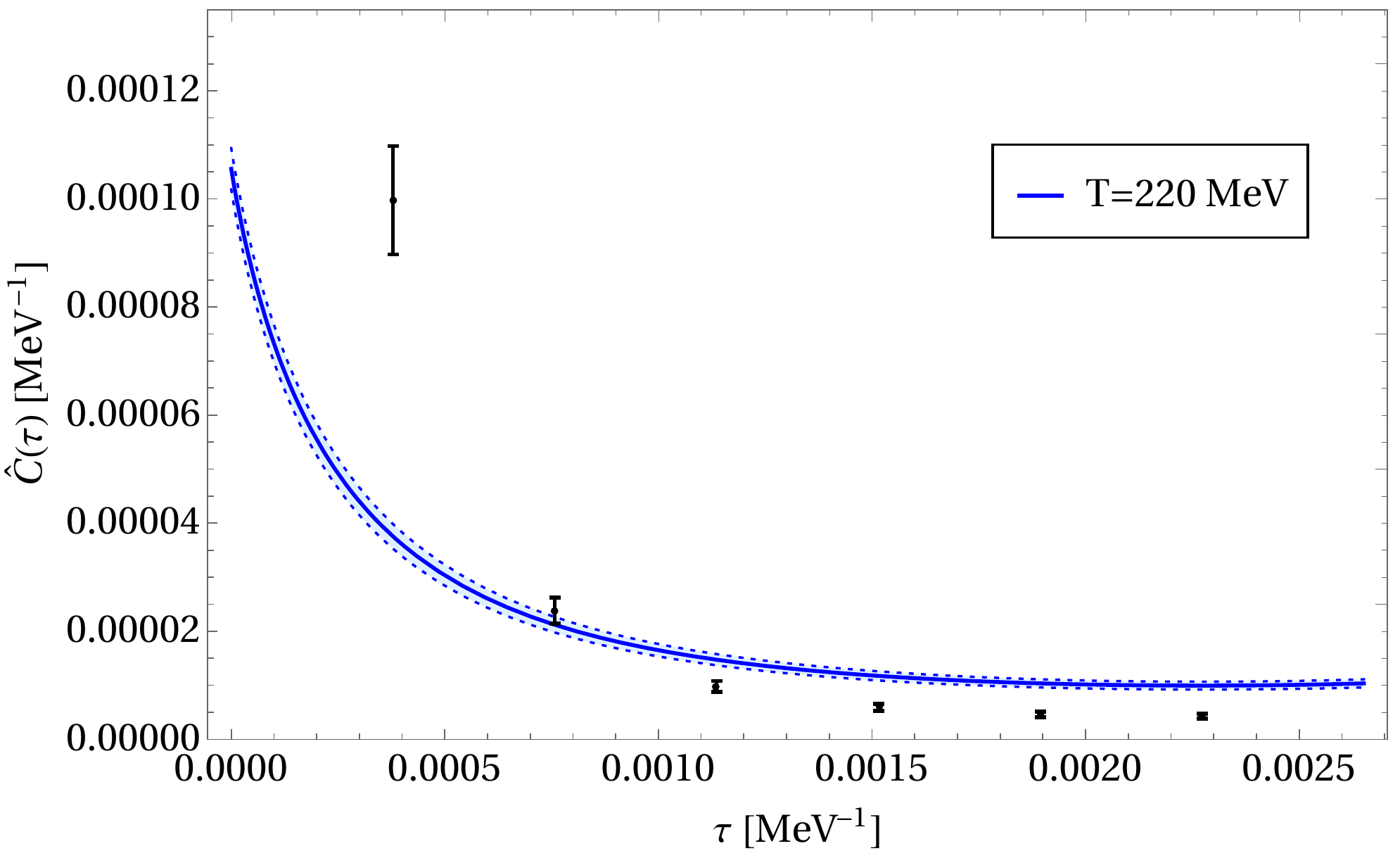}
\caption[]{Left: Temporal correlation function predicted by the spectral function \eq(\ref{eq:spec1}), 
\fig\ref{fig:spec} (red band), compared
to the full lattice data  from \fig\ref{fig:chi_spin2} \cite{Rohrhofer:2019qal}.
Right: The corresponding prediction based on a Breit-Wigner ansatz, \eq(\ref{eq:spec2}). }
\label{fig:temp}
\end{figure}
Since
we neglected the continuum contribution from \eq(\ref{eq:decomp}), it is crucial to perform a quality check. This is done
in \fig\ref{fig:temp} (left), where we predict the temporal correlator $C^\tau_{PS}$ using our spectral function from the spatial correlator 
at $T=220$ MeV, and compare with the lattice result from \fig\ref{fig:chi_spin2}. Excellent quantitative agreement is found except
for very short distances, which is due to the neglected higher excited states in the description of the spatial correlator.
For higher temperatures we expect the quality of the prediction to deteriorate, as in this case the 
neglected continuum part
$D_{c,\beta}$ should play an increasing role.

For comparison, we have also tried a Breit-Wigner ansatz commonly associated with perturbative plasma excitations, 
\begin{equation}
\rho_{PS}^{BW}(\omega,\mathbf{p}=0) =
\frac{4\alpha_\pi \omega\Gamma_{\pi}}{(\omega^2-m_\pi^2-\Gamma_\pi^2)^2+4\omega^2\Gamma_\pi^2}+
\frac{4\alpha_\pi^*\omega\Gamma_{\pi^*}}{(\omega^2-m_{\pi^*}^2-\Gamma_{\pi^*}^2)^2+4\omega^2\Gamma_{\pi^*}^2}\;.
\label{eq:spec2}
\end{equation}
This ansatz can be fitted equally well to the spatial correlator at $T=220$ MeV, but in this case the predicted temporal correlator 
is not compatible with the data, \fig\ref{fig:temp} (right).  

\section{The QCD phase diagram}

Having established an effectively chiral spin symmetric temperature window at zero density with non-perturbative dynamics, 
the question is what
happens at finite baryon chemical potential. This adds $\mu_B/3\;\bar{\psi}\gamma_0\psi$ to the 
effective Lagrangian, and since the generators of chiral spin commute with $\gamma_0$, \eq(\ref{eq:comm}), an approximate chiral
spin symmetry at zero density must continue to $\mu_B\neq 0$. At least for $\mu_B/T\lsi 3$ we can then infer what
happens to the chiral spin symmetric band. 
\begin{figure}[t]
\vspace*{-0.5cm}
\centering
\includegraphics[width=0.45\textwidth]{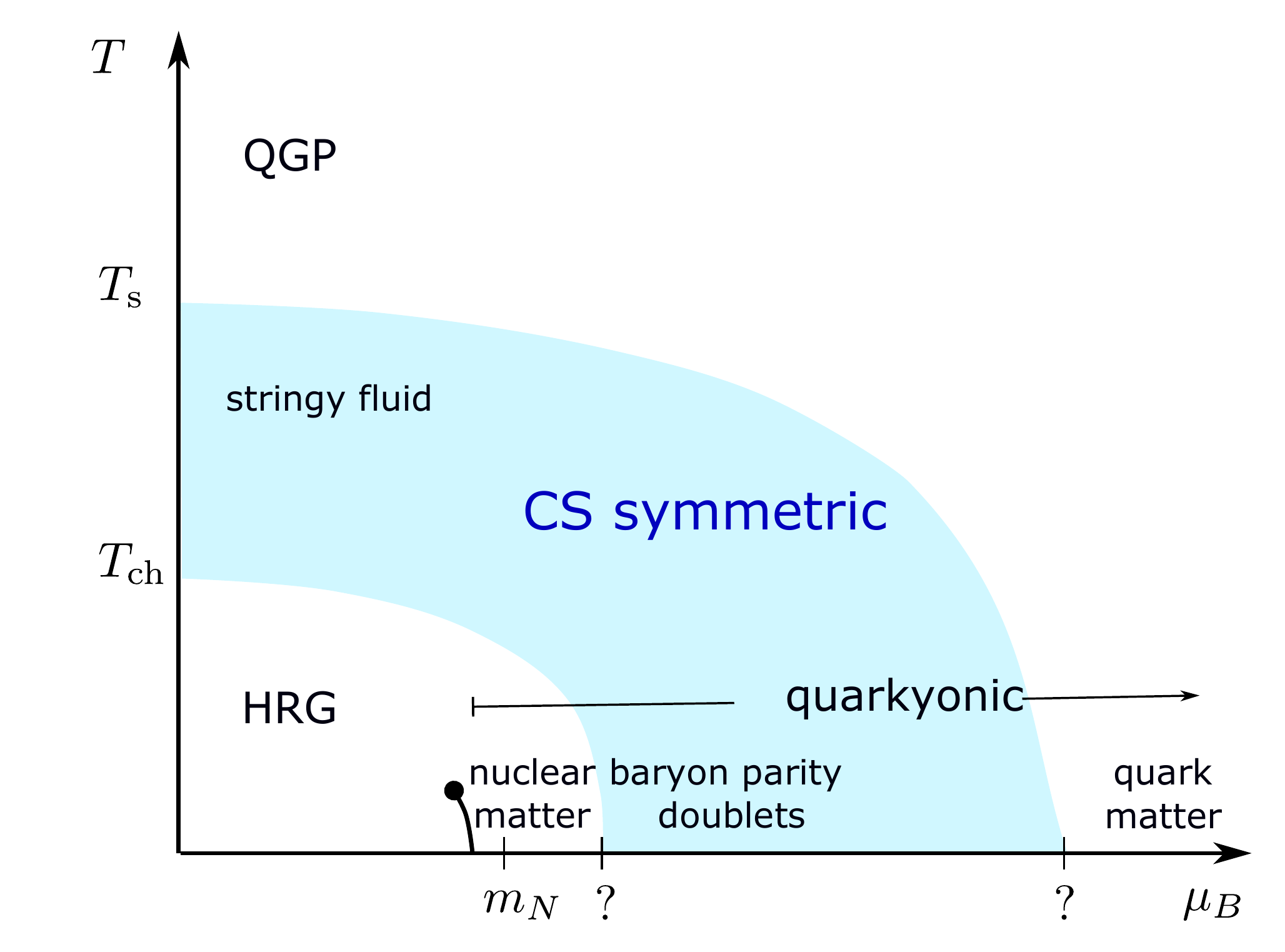}
\includegraphics[width=0.45\textwidth]{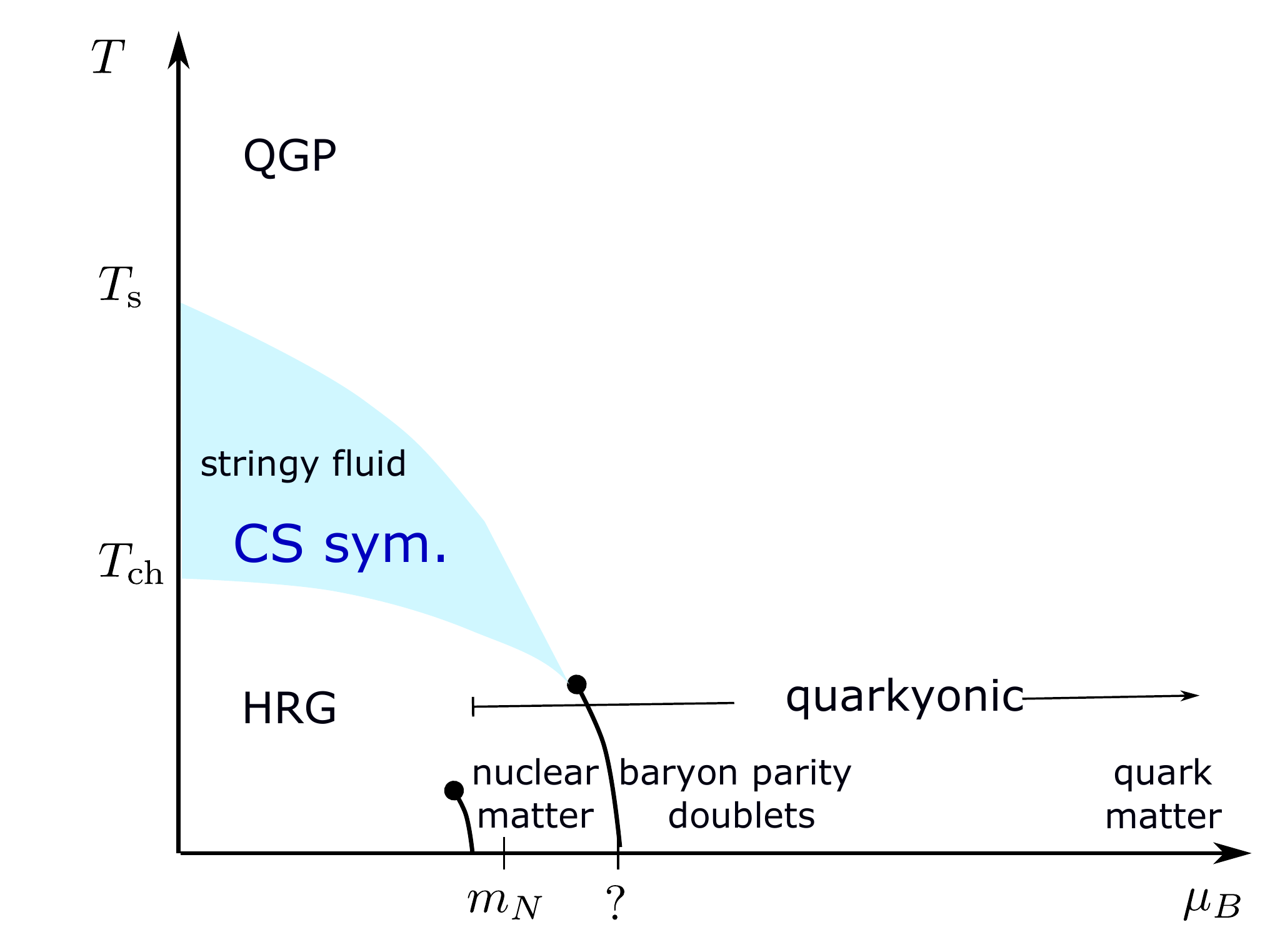}
\caption[]{Possibilities for the  QCD phase diagram with a chiral spin and $SU(4)$-symmetric band.}
\label{fig:pd}
\end{figure}

Since full chiral symmetry restoration is necessary for chiral spin symmetry, its lower 
boundary $T_\mathrm{ch}(\mu) \gsi T_\mathrm{pc}(\mu)$. 
In Sec.~\ref{sec:masses} we identified the upper crossover $T_\mathrm{s}$ by the bending of the screening masses, 
which marks the screening of the colour-electric interactions and the onset of perturbative behaviour. 
Picking the vector meson screening radius at its bend to define the screening temperature $T_\mathrm{s}$,
\beq
r_V^{-1}(\mu_B=0, T_\mathrm{s})\equiv m_V(\mu_B=0, T_\mathrm{s})=C_0T_\mathrm{s}\;,
\eeq
we can use the Taylor expanded screening mass to deduce the line of constant $r_V^{-1}$,
\beq
\frac{m_V(\mu_B)}{T}=C_0+C_2\left(\frac{\mu_B}{T}\right)^2+\ldots\quad \Rightarrow\quad
\label{eq:scr_mu}
\frac{dT_\mathrm{s}}{d\mu_B}=-\frac{2C_2}{C_0}\frac{\mu_B}{T}-\frac{2C_2^2}{C_0^2}\left(\frac{\mu_B}{T}\right)^3+\ldots\;.
\eeq 
Since $C_2>0$  \cite{Hart:2000ef,Pushkina:2004wa}, the upper crossover line bends downwards, as indicated in \fig\ref{fig:pd}.

As chemical potential increases, further details of the phase diagram remain unknown,
and several options for the continuation of the chiral spin symmetric band are possible. 
In the cold and dense regime, baryon parity doublet matter is consistent with chiral spin symmetry, provided it is 
decoupled from $\pi,\sigma$ to leading order, otherwise it is only chirally symmetric. Similarly, quarkyonic matter 
with a chirally symmetric confined regime~\cite{McLerran:2007qj,Philipsen:2019qqm} may also be chiral spin symmetric, 
as discussed in \cite{Glozman:2022lda}. 
This is independent of the question whether
or not there  is a non-analytic 
chiral phase transition. Two possibilities (there are more) for the phase diagram are shown in \fig\ref{fig:pd}. 

\section{Conclusions and outlook}

For QCD at the physical point at zero density, there appear to exist three temperature regimes with different effective 
symmetries, separated by two crossovers. The low temperature hadronic regime with broken chiral symmetry is, somewhere
above the chiral crossover $T_\mathrm{pc}$, followed by a regime with an approximate $SU(4)$ symmetry, which features
non-perturbative dynamics as shown by screening masses, and resonance-like pion states. 
This is consistent with a dominant colour-electric quark-gluon interaction, as suggested by the effective symmetry.
At a second crossover around
$\sim 3T_\mathrm{pc}$, the symmetry is reduced to the expected chiral symmetry, screening masses follow perturbative 
behaviour and the resonance-like pion peaks have disappeared. 
We have argued by symmetry and the known behaviour of screening masses, that this chiral spin symmetric band 
should continue to finite baryon density and bend downwards, with several possibilities for the QCD phase diagram.

There is now an entire range of feasible lattice projects to clarify the range and properties of the chiral spin symmetric band. 
In particular, the pseudo-critical
$T_\mathrm{s}(\mu)$ can be mapped out just as far as $T_\mathrm{pc}(\mu)$, using lattice methods such as imaginary
chemical potential, Taylor expansion or reweighting. To obtain further insight into the dynamics, spectral functions should
be studied in other quantum number channels as well. To this end, we have demonstrated that much tighter
constraints are obtained once both spatial \textit{and} temporal correlators are available with good precision. 

\acknowledgments
O.P.~and P.L.~acknowledge support by the Deutsche Forschungsgemeinschaft (DFG) through the 
grant CRC-TR 211 ``Strong-interaction matter under extreme conditions''. O.P.~further acknowledges support
 by the State of Hesse within the Research Cluster ELEMENTS (Project ID 500/10.006).

\bibliographystyle{JHEP}
\bibliography{references}

\end{document}